White Paper

on

# IMPROVEMENTS TO THE NASA RESEARCH AND ANALYSIS PROPOSAL AND REVIEW SYSTEM

| | |
|---|---|
| Paul Byrne | NC STATE UNIVERSITY \| paul.byrne@ncsu.edu |
| Christina Richey | JET PROPULSION LABORATORY, CALTECH |
| Julie Castillo-Rogez | JET PROPULSION LABORATORY, CALTECH |
| Mark Sykes | PLANETARY SCIENCE INSTITUTE |

**Key Recommendations**
1. The SMD must urgently and substantially increase the funds available for R&A programs to bolster selection rates and ensure a sustainable future for the planetary science workforce
2. NASA should avoid, to the greatest extent possible, the changing of previously annual programs to a biennial cadence, and the modifying of anticipated budget levels and numbers of new starts for programs once those programs have been announced
3. The SMD should allow resubmitted proposals above some merit threshold to address earlier shortcomings, in the manner of peer review for academic manuscripts, and consider the value of having some continuity in reviewers over successive years
4. NASA should consider a two-stage review process in which detailed budget justifications are sought only for those proposals that pass some selectability threshold following peer review
5. NASA should compensate external reviewers
6. The SMD should review whether the criteria for assessing conflicts of interest are overly broad, especially for programs where candidate reviewer pools are small
7. NASA should minimize the instances of hybrid panels until more experience is gained in how to balance the influence of remote and in-person reviewers
8. NASA should clearly and effectively communicate the results of dual-anonymous peer review to the community, together with a transparent rationale for whether to discontinue or expand it
9. Continued implicit bias training is crucial to ensuring that the harm of reviewer and PO biases is minimized, and such training should be expanded to external reviewers
10. NASA should permit review panels to commend proposals planning to train early-career researchers
11. SMD should explicitly fund FINESST as a stand-alone program, and offer additional sources of continued support for junior scientists

## 1. Introduction

NASA's Research and Analysis (R&A) programs represent a major source of support for the U.S. planetary science community. The results of an American Astronomical Society (AAS) Division for Planetary Sciences (DPS) workforce survey earlier in 2020 found that the majority of planetary scientists are supported by NASA R&A funds (Hendrix et al., 2020). Indeed, almost all respondents to the workforce survey had submitted as principal investigator (PI) at least one research proposal to a NASA Research Opportunities in Space and Earth Science (ROSES) program, and 34% of non-faculty respondents receive the *majority* of their funds from R&A programs (Hendrix et al., 2020). The role of these programs in maintaining planetary research in the United States cannot be overstated.

But the health of our community is threatened by inadequacies and inefficiencies in the proposal preparation and review process, which together increase the burden on researchers to secure funding and, ultimately, risks their leaving the field. Unsustainably low selection rates because of too little money is perhaps the most obvious issue we face—documented more fully in other white papers including those by Castillo-Rogez et al. (2020) and Sykes et al. (2020)—but other problems persist. For example, the review process itself offers no reliable way for proposers to improve their selection success when resubmitting proposals, substantial work hours are involved in preparing proposals, the vast majority of which will not be selected, and the factors underpinning "programmatic balance" when making selections for a given R&A program are often not clear to the community.

The NASA Science Mission Directorate (SMD) has recently announced steps to improve the R&A proposal and review process, including trialing dual-anonymous peer review (DAPR) (see the white papers by Aloisi and Reid (2020) and Radebaugh et al. (2020)), and considering the adoption of a "no due date" policy for some programs (see the presentation by Dr. S. Rinehart to the NASA Planetary Science Advisory Committee (PAC) in August 2020). These steps are encouraging, but there is more to do. In this white paper, we address some remaining issues and offer solutions that NASA can implement to reduce the burden of both proposers and reviewers, enable a reduction in biases for reviewers and program officers (POs), and help sustain a healthy research community generally.

## 2. R&A Selection Rates
### *2.1. Issues*

Selection rates for key NASA R&A programs have stabilized at about 20% for the past several years (***Figure 1***), a level considerably below that recommended by the 2013–2022 "Visions and Voyages for Planetary Science" Decadal Survey report, which stated (emphases our own):

> *"Over the seven fiscal years 2003–2009, on average* **37 percent** *of the grant proposals submitted to an average of 18 or 19 programs in NASA's Planetary Science Division were supported.* The success ratio is lower than desirable… *The [Decadal Survey] committee stro***ngly encourages NASA to find ways to increase average grant sizes and reduce the number of proposals that must be written, submitted, and reviewed by the community***."*

But in the presentation to the NASA PAC in August 2020, NASA reported that selection rates for arguably its flagship program, Solar System Workings, have been at an approximately constant rate of 20% for ROSES years 2014 to 2018, before dropping precipitously to 11% in ROSES-19 (***Figure 2***). Moreover, the ~20% selection rates were facilitated by borrowing against future years, and by canceling the ROSES-18 call for the Planetary Science and Technology from Analog Research (PSTAR) program. This practice of artificially inflating selection rates by borrowing from future years is clearly not sustainable.

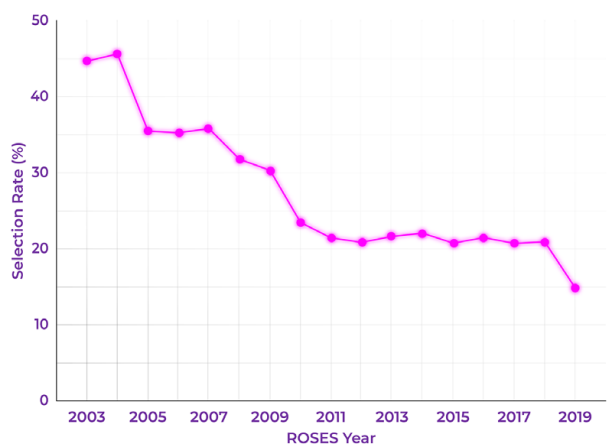

**Figure 1.** *Selection rates for NASA R&A programs from ROSES years 2003 through 2019. For most of the last decade, selection rates have remains relatively stable at ~20%, but dropped substantially for ROSES-19. Data compiled by co-author Sykes.*

Low selection rates have numerous, negative consequences for both researchers *and* NASA. For example, ensuring successful funding when selection rates are proportionately low requires researchers to spend proportionately more time writing proposals. To increase their chances of having a given proposal selected, researchers are writing more complex proposals with more Co-Is, translating to more *expensive* proposals—exactly as NASA has seen, with average requested budgets for the SSW program rising by ~40% from ROSES-14 through ROSES-19 (***Figure 2***). And decisions to postpone or cancel programs are detrimental to researchers planning near- to medium-term research priorities and goals, especially early-career, untenured, and soft-money scientists. Canceling the PSTAR program in ROSES-18 and the announcement in earlier 2020 that PSTAR will transition to a biennial call, as well as downward revisions of expected funding amounts and selection percentages for open calls—as for the Planetary Instrument Concepts for the Advancement of Solar System Observations (PICASSO) call in ROSES-20, which saw a ~14% reduction in its budget—pose further difficulties to a community severely under pressure.

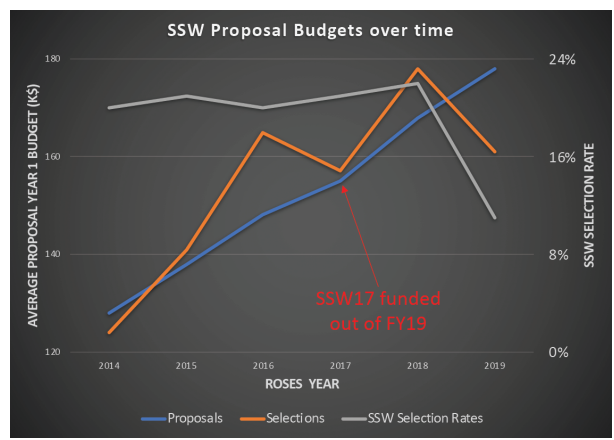

**Figure 2.** *Selection rates for NASA's Solar System Workings (SSW) program from ROSES-14 through ROSES-19 (grey line). The average cost of submitted and selected proposals are shown for that same time period by blue and orange lines, respectively. This plot is taken from the presentation given by Planetary Science R&A Lead Dr. S. Rinehart to the NASA PAC meeting in August 2020.*

*2.2. Solutions*
The solution to these problems is simple: NASA must fund R&A programs at substantially higher levels than at present, at a rate suitable to ensure that *at least* 37% of proposals are selected—the number identified in the 2013–2022 Decadal Survey report. We emphasize, however, that such a percentage should be regarded as a *minimum*. The healthier the average selection rates for R&A programs, the more effectively and sustainably our community will be able to carry out a primary strategic goal of NASA: expand human knowledge through new scientific discoveries. We refer the reader to the white paper by Castillo-Rogez et al. (2020) for further discussion of what R&A selection rates might appropriately constitute a "healthy" funding level.

*2.3. R&A Selection Rate Recommendations*
- **The SMD must urgently and substantially increase the funds available for R&A programs to bolster selection rates and ensure a sustainable future for the planetary science workforce in the United States.**
- **NASA should avoid, to the greatest extent possible, the changing of previously annual programs to a biennial cadence, and the modifying of anticipated budget levels and numbers of new starts for programs once those programs have been announced.**

## 3. Proposal Resubmissions
*3.1. Issues*
A recurring issue many planetary researchers face is inconsistency in review panel feedback in successive years, meaning that even if a scientist receives useful information on their proposal from a panel one year, that scientist can have little expectation that a revised proposal will fare any better the next year. NASA POs publicly state that proposals are treated as new each time they are submitted, and the make-up of review panels understandably changes from year to year.

The lack of any consistency between successive peer review assessments of R&A proposals promotes the perception of reviews being stochastic, and contrasts strongly with established procedures for the review of academic journal articles. Few scientists would welcome one set of reviewer comments on an initial draft of a submitted manuscript, only to face potentially a whole new set of criticisms upon revision! This inconsistency represents a material problem, set against the ever-present backdrop of unsustainably low selection rates. It is in NASA's interest to offer useful guidance and consistent assessments to proposers from one year to the next.

NASA PO Dr. H. Throop carried out a "microstudy" with Cassini and New Frontiers Data Analysis Program proposals in ROSES-18 and -19, finding that of 29 resubmitted proposals the average change in "merit" score was +0.5—a 10% improvement in absolute score (**Figure 3**). But at least one proposal saw no change in score, and the scores of seven proposals (~25%) *decreased*. CDAP in particular is an established and stable program; the results of a similar study for SSW may well be very different.

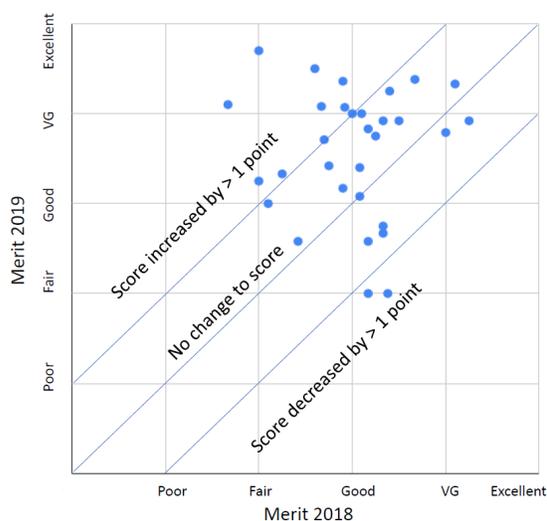

**Figure 3.** *The results of a microstudy by NASA PO Dr. H. Throop for proposals resubmitted to the ROSES-18 and ROSES-19 CDAP and NFDAP programs. This plot is taken from the presentation given by PO Throop to the NASA Outer Planets Assessment Group meeting in February 2020.*

*3.2. Solutions*
Several options are available to NASA to address these problems. In the 1980s and 1990s, it was expected that new awardees of a given program would be available to serve on the review panel of that program for two or three years or, alternatively (and to avoid reinforcing panel member biases), provide external reviews over that same period. For resubmitted proposals, reviewers had access to prior year versions of the proposal in addition to its reviews. Indeed, there is modern precedence for tying an award to future review service: the Planetary Data Archiving, Restoration, and Tools (PDART) program at present requires PIs of selected proposals to provide, as part of the terms of their award, peer reviews for two other PDART-supported map projects.

Further, NASA could permit proposers resubmitting a previously declined proposal to directly address, in a specific section, the merit weakness of the predecessor proposal—exactly as is the case for journal manuscripts. To reduce the workload of POs and panelists, a merit score threshold (e.g., ≥"Good") might apply to such proposals. Highlighting resubmitted proposals as exactly that, in contrast to the current practice of appraising every submission every year as new, would enable successive review panels to check whether revised proposals have materially addressed the previous year's feedback. In parallel, POs should monitor which proposals are resubmissions, work with the review panels to assess those revised proposals, and above all ensure that successive panels do not return contradictory reviews from one year to the next on a given proposal that clearly enumerates how the proposers have responded to an earlier round of criticism.

### *3.3. Recommendation*
- **The SMD should allow resubmitted proposals above some merit threshold to address earlier shortcomings, in the manner of peer review for academic manuscripts, and consider the value of having some continuity in reviewers over successive years.**

## 4. Preparing Proposal Budgetary Information
### *4.1. Issues*
Since ROSES-16, proposals have been required to include only limited budgetary data in the material seen by reviewers, supplying additional detail in separate "Total Budget" documents. The rationale for this change is sound: proposers cannot influence their institution's negotiated Cost Rate Agreement, and thus the total cost of their proposal.

But the gathering of proposal budgetary documents can be onerous, especially for projects with numerous Co-Is, and for those PIs at smaller institutions without strong backroom support (such as a well-staffed proposal preawards office at an R1 university). Considerable effort is needed to ensure compliance with proposal formatting and content (especially regarding material outside the science/technical/management section, notably the budget justification section)—and this effort is presently required for all proposals, even though >80% will not be selected, and these details of a proposal may differ substantially from year to year.

Further, anecdotal experience of having served on review panels indicates that panelists are asked to assess the "Cost" component of a proposal in broad terms only, e.g., by assigning scores of "1", "2," or "3," with little scope for making specific findings despite being asked to assess current and pending information, assess justifications and detailed breakdowns for direct costs such as travel and related expenses, etc.

### *4.2. Solutions*
The planned implementation of dual-anonymous peer review offers NASA the opportunity to experiment with an approach in which proposals are at first assessed only for Merit and Relevance—with only the budgetary information of those proposals passing some threshold (e.g., ≥"Good," or some other measure that renders a proposal "selectable") being then considered by review panels, or even by the POs alone after the formal review ends. Proposals would still list levels of effort for project personnel, but panels should not be tasked with analyzing anything beyond this information, which could be made a component of the "Merit" score.

Under this approach, costs would be estimated by proposers (i.e., total direct and indirect costs, encompassing work effort, travel, publications, etc.), quoted by project year (e.g., "Y1 total cost = $150,000"). If their submissions are deemed "selectable," proposers would then provide detailed budgetary information consistent with those earlier estimates to aid POs in making final selections—avoiding the need for proposers to spend considerable work hours collating financial information for proposals that, statistically, are unlikely to be selected. Further, the time spent by review panels would also be reduced, with commensurate reductions in expenses associated with honoraria and travel, and an improved ratio between work hours spent reviewing and the number of selected proposals.

### *4.3. Recommendation*
- **NASA should consider a two-stage review process in which detailed budget justifications are sought only for those proposals that pass some selectability threshold following peer review.**

## 5. Peer Review Mechanisms
### *5.1. Issues*
A critical component of NASA's R&A program is the peer review performed by members of the community—a reasonable and necessary service, but one that carries its own challenges. For example, reviews take time, both during the time a panel meets (whether in person or virtually), and beforehand as panelists prepare preliminary findings. External reviews, which can provide important supplementary insight into proposals especially if a panel lacks a particular expertise, are not compensated even though they, too, take time. The anecdotal experience of this white paper's authors is that many, and perhaps most, external review requests are not fulfilled. And rules governing conflicts of interest can also act to impede the review process, substantially decreasing the pool of eligible reviewers most acutely for major programs such as SSW.

In-person panels can be helpful for some people, such as those scientists who face challenges in securing uninterrupted blocks of time at home, and for panel members in different time zones. But virtual panels may suit others better, such as those researchers with disabilities for whom travel is difficult, and responses to the move to exclusively online panels because of the COVID-19 pandemic have generally been favorable. Nonetheless, panels with both physically present and virtual members can alter the dynamic of the group in favor of those in the room together.

### *5.2. Solutions*
Providing an honorarium or other financial recompense to eligible planetary scientists would help increase the uptake of external review requests. To be clear, that remuneration should be in *addition* to the increased funding for R&A programs that we advocate for here—there is no sense in attaining more money for scientific research only to spend it on reviewing proposals for such research. (Compensation could also include augmentation of an existing award of a reviewer.)

Assessing which programs can support virtual review panels, and those that would benefit from in-person meetings, has the potential to increase the eligible reviewer pool by enabling those scientists to participate in panels who would otherwise decline to travel. Equally, even in the post-COVID-19 world in-person reviews may be preferential to some scientists and POs. Where possible, however, mixed-mode reviews should be avoided, to ensure as level a playing field as possible for both reviewers and reviewed proposals . Reviewing the circumstances in which conflicts of interest may pose a real risk to the fair evaluation of a proposal, instead of hindering the effective review of proposals for which subject-matter experts may be in short supply, would further enhance the quality of the review process.

### *5.3. Recommendations*
- **NASA should compensate external reviewers.**
- **The SMD should review whether the criteria for assessing conflicts of interest are overly broad, especially for programs where candidate reviewer pools are small.**

- **NASA should minimize the instances of hybrid panels until more experience is gained in how to balance the influence of remote and in-person reviewers.**

## 6. Biases in Peer Review
### *6.1. Issues*
Numerous issues regarding Diversity, Equity, Inclusion, and Accessibility (DEIA) have been raised in the planetary science community in recent years—with several DEIA-focused white papers led by Rathbun et al. (2020), Richey et al. (2020), Rivera-Valentín et al. (2020), Scalice et al. (2020), Strauss et al. (2020), and others to which we refer the reader. But biases of all kinds toward researchers pervade our community, including in the actual mechanics of proposal peer review. It is therefore critical that we work to recognize and mitigate these biases as much as possible, for a healthy R&A-supported community is not only one that is sufficiently funded, but is one that includes sufficient work to lower the barriers faced by underrepresented and otherwise marginalized planetary researchers.

For example, biases (including unconscious biases) held by reviewers can unfairly target proposers from minority groups and/or specific institutions. Similarly, the biases of R&A personnel can affect the specific selection decisions made by program officers and selecting officials, especially because there is little transparency in those decisions that underpin "programmatic balance" within SMD after review panels formalize their findings for NASA. Working to ensure such balance between and within R&A programs makes sense, but it is critical that the selection decisions therein are made without the undue influence of bias toward a proposer's personal characteristics, institution, etc.

### *6.2. Solutions*
As we note above, we are encouraged by NASA's decision to consider dual-anonymous peer review (DAPR) for some R&A programs—although we note that PDS's trial program, Habitable Worlds, is not solicited in ROSES-21 for budgetary reasons and so the results from this first DAPR experiment will not be available for some time. (Several data analysis programs (DAPs) will feature DAPR for ROSES-21.) Nevertheless, it is important that the results of this trial be clearly and rapidly communicated to the community, together with the reasoning behind any decision to expand, or cancel, DAPR practices.

But DAPR will not fully mitigate the risks to fair and equal peer review of personal biases. Therefore, NASA should continue the welcome approach of providing implicit bias training for panelists at the start of a review, but this training *should also be expanded to external reviewers*. Such training should be mandatory after a scientist accepts a review assignment but before they can access proposal(s); compensating external reviewers (see above) would help with this requirement.

Additionally, regular implicit bias training for R&A personnel, and an explicit anti-bias focus in selection documents, will help ensure that those staff members making selection decisions are not being influenced by their own biases. Indeed, NASA's Astrophysics Division features an approach whereby POs act as levelers in review panel discussions, staying in one assigned room for the duration of the review. Such an approach could be trialed by PSD to both further identify and mitigate biases.

### *6.3. Recommendations*
- **NASA should clearly and effectively communicate the results of DAPR to the community, together with a transparent rationale for whether to discontinue or expand the practice.**
- **Continued implicit bias training is crucial to ensuring that the harm of reviewer *and* PO biases is minimized, and such training should be expanded to external reviewers.**

## 7. Supporting the Next Generations of Planetary Scientists
### *7.1. Issues*
NASA does not have an explicit education remit, although its education and public outreach (EPO) activities have been considerably reduced over the past decade. Nonetheless, NASA has few peers in its ability to influence and inspire younger generations of students to pursue planetary- and space-related training and careers. It is these people who will become the next generation of planetary scientists, and we must do everything we can to support them.

Yet there is no explicit requirement for, nor means of rewarding, the inclusion of undergraduate, graduate, or postdoctoral researchers on NASA R&A proposals, even when those proposals would materially train such early-career researchers in important research skills. Although SMD supports annual calls for the Future Investigators in NASA Earth and Space Science and Technology (FINESST) program, FINESST selections are funded from thematically relevant mainstay R&A programs, such as SSW. Fluctuations in funding levels of these programs therefore can have commensurate effects on FINESST selections, affecting the types of research early-career scientists are able to pursue.

### *7.2. Solutions*

The ability to afford review panels the opportunity to include as a merit strength finding—even at the level of a minor strength—an explicit plan in a given proposal to train student or postdoctoral researchers would represent a meaningful demonstration by NASA that the training of early-career scientists is valued. That is *not* the say that such student or postdoctoral researchers are or should be necessarily bound for an academic or research career: particularly with selection rates as low as they are, *there is too little money for the size of the community as it is*.

But equally, academic or research careers are not the only course open to early-career scientists; much more is needed to normalize non-academic career paths in planetary science, as Frank et al. (2020) discuss in more detail in their white paper. And, importantly, proposals that do *not* include junior scientists should not be discriminated against: the merit of training future planetary scientists is not a zero-sum game, and not all projects can support, nor have access to, students or postdoctoral scholars.

NASA can take additional steps to visibly support, and secure, the training of the next generation of planetary scientists. Supporting FINESST as an independent program, free of the influence of fluctuating funding levels for other programs , would enable Future Investigator candidates and their PIs to propose projects that extend beyond the scope and remit of existing, mainstream R&A calls. Providing additional opportunities for junior researchers would further help, such as sponsoring travel awards not only for assessment/advisory group meetings, but for conferences and workshops, too. The recently announced plan to involve early-career researchers at mission team meetings is a positive step in this direction, and one we encourage NASA to develop.

### *7.3. Recommendations*

- **NASA should enable and encourage review panels to commend proposals for including training for early-career researchers.**
- **SMD should explicitly fund FINESST as a stand-alone program, and offer other sources of continued support for junior scientists.**